\documentclass[twocolumn,aps,showkeys,showpacs,amsmath,amssymb]{revtex4}

\usepackage{graphicx}
\usepackage{bm}

\begin{document}

\title{Large LFMR observed in twinned $\rm La_{2/3}Ca_{1/3}MnO_{3}$ thin films epitaxially grown on YSZ-buffered SOI substrate}

\author{J. Li}

\author{P. Wang}\author{J. Y. Xiang}\author{X. H. Zhu}\author{W. Peng}\author{Y. F. Chen}\author{D. N. Zheng}
\affiliation{Superconductivity Division, Beijing National Laboratory for Condensed Matter Physics, Institute of
Physics, Chinese Academy of Sciences, Beijing 100080, China}

\author{Z. W. Li}
\affiliation{Temasek Laboratories, National University of Singapore, Engineering Drive 3, Singapore 119260}

\date{\today}

\begin{abstract}
$\rm La_{2/3}Ca_{1/3}MnO_{3}$ thin films have been grown on yttria-stabilized zirconia (YSZ) buffered
silicon-on-insulator (SOI) substrate by the pulsed laser deposition technique. While full cube-on-cube epitaxy was
achieved for the YSZ layer, the top manganite layer was multi-domain-oriented, with a coexistence of cube-on-cube
and cube-on-diagonal epitaxy. Due to a combined influence from the magnetocrystalline anisotropy and the
magnetoelastic anisotropy, in zero field the local spin orientation varies across the twin boundaries. As a
result, a quite large low-field magnetoresistance (LFMR) based on spin-dependent tunnelling was observed. The film
shows a resistance change of $\sim$20\% in a magnetic field $<$ 1000 Oe at 50 K, which is promising for real
applications.
\end{abstract}


\maketitle

Manganite materials with perovskite structure have attracted considerable attention during the past decade, since
a \lq colossal magnetoresistance\rq~occurs around their Curie temperature $T_C$ \cite{Jin1,Jin2}. Potential
applications such as new-generation magnetic sensors or reading heads are ardently anticipated. Numerous efforts
have been made to prepare manganite thin films on single-crystalline substrates of SrTiO$_3$, LaAlO$_3$, MgO, and
the yttria-stabilized zirconia (YSZ). However, in order to realize an integration with the existing silicon
manufacturing technology, it is practically important to grow high quality manganite thin films on silicon wafers.
Nevertheless a silicon substrate is unfavorable for a full epitaxy growth, due to not only its large lattice and
thermal expansion mismatch with the manganites, but also its easily oxidized surface and severe chemical inter
diffusion. Therefore up to now high quality manganite thin films were achieved only through heteroepitaxial growth
with several transition layers \cite{Wong,Fontcuberta,Kim}. Lately, in the semiconductor industry a new
technology, SOI (silicon on insulator), has become prevailing \cite{Shahidi}. It has been suggested to take
advantage of the buried polycrystalline silicon dioxide (SiO$_2$) in the wafer as a compliant layer to relax the
strain between the films and the substrate, then high quality crack-free thin films can be obtained on
single-layer-buffered SOI substrate \cite{Kang,Wang}. In this letter, we report the growth of $\rm
La_{2/3}Ca_{1/3}MnO_{3}$ (LCMO) thin films on YSZ-buffered SOI substrate. A unique crystallographic domain
structure with a high density of 45$^\circ$ twins were observed. As a result the LFMR is enhanced due to the
spin-dependent tunnelling across the twin boundaries.

We adopted commercial p-type SIMOX (100)SOI wafers provided by Shanghai Simgui. Prior to the film deposition, the
SOI substrate was ultrasonically cleaned in diluted phosphoric acid (4\%), deionized water, alcohol and acetone
sequentially to remove contaminations from its surface, but leave the topmost native SiO$_2$ layer untouched. The
films were grown by the pulsed laser deposition technique using a KrF excimer laser. The YSZ target was a single
crystal, while the LCMO target was a stoichiometric ceramic. The chamber was evacuated down to $5\times 10^{-4}$
Pa before the heating started in order to avoid further surface oxidation. The YSZ buffer layer was grown in the
background vacuum at a substrate temperature ($T_S$) of 820$^\circ$C and a laser repeat rate of 7 Hz. The LCMO
thin film was then deposited \textit{in situ} in an oxygen pressure of 40 Pa on YSZ at $T_S$ of 860$^\circ$C and a
repeat rate of 2 Hz. High purity oxygen was injected in immediately after the film deposition, then the films were
cooled down in 1 atm oxygen at a ramp rate of 40$^\circ$C/min to room temperature. The film thickness was around
100 nm for YSZ and 150 nm for LCMO.

\begin{figure}
\begin{center}
\includegraphics[width=0.60\columnwidth]{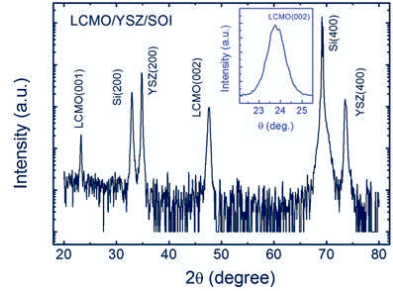}
\caption{\label{XRD}XRD $\theta-2\theta$ scan of the LCMO thin film grown on a YSZ-buffered (100)SOI substrate.
Inset is the rocking curve of LCMO(002).}
\end{center}
\end{figure}

\begin{figure}
\begin{center}
\includegraphics[width=0.9\columnwidth]{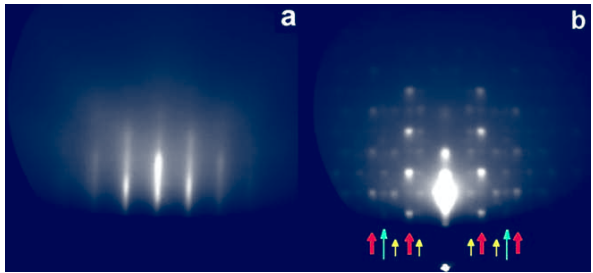}
\caption{\label{RHEED}RHEED patterns for the YSZ buffer layer (a), and the top LCMO thin film (b). The incident
electron beam (25 kV) was along Si[100].}
\end{center}
\end{figure}

Growth of the YSZ layer was monitored \textit{in situ} by a real-time RHEED, while the in-plane crystallinity and
the morphology of the LCMO layer was checked by RHEED at room temperature after the film deposition. The film
out-of-plane orientation and crystallinity were characterized by x-ray diffraction (XRD) $\theta$-$2\theta$ scan
and $\omega$-scan. The electrical resistivity ($\rho$) and the magnetoresistance (MR) of the films were measured
by using a standard four-probe technique. An MPMS£­5 superconducting quantum interference device (SQUID) was
employed to generate uniform magnetic field for both MR and magnetization measurements.

Figure~\ref{XRD} shows a typical XRD $\theta$-$2\theta$ spectrum of the LCMO/YSZ/SOI heterostructure. In the
figure, the LCMO thin film is indexed as a pseudocubic structure. Besides the reflections from Si substrate, only
the (200) and (400) of YSZ and the (001) and (002) of LCMO are observed, revealing the full \textit{c}-axis
texture of the film. Here Si(200) appears due to the long time heating during the film deposition. The
out-of-plane lattice parameters derived are 5.15 \AA~for YSZ and 3.82 \AA~for LCMO. The inset is the rocking curve
of the LCMO(002). Its full width at half maximum (FWHM) is about 0.95$^\circ$, indicating a small mosaic spread
along \textit{c}-axis in the film.

The in-plane orientations of the bilayer were determined by RHEED observations. The incident electron beam was
along Si[100]. After the first five to ten laser pulses, we noted that the RHEED pattern of SOI substrate faded
out, implying that the first several YSZ atomic layers are of poor crystallinity. This most probably originates
from the native amorphous SiO$_2$ layer on the SOI surface. Nevertheless, 30 seconds later, a blur and thick
streaklike diffraction pattern of YSZ[100] appeared. With increase of the deposition time and thus increase of the
film thickness, the intensity and sharpness of the diffraction pattern gradually increased, suggesting a recover
of the YSZ crystallinity. Fig. \ref{RHEED}(a) was recorded after the YSZ film deposition. The very clear and sharp
streaks strongly suggest perfect in-plane epitaxy of the film, as well as its 2D-like atomic smooth surface
\cite{Wang2}. The LCMO thin film grown on YSZ, however, shows clear 3D characteristics, as seen in
Fig.~\ref{RHEED}(b). The roundish instead of streaklike diffraction spots reveal that the surface of the LCMO film
is rather rough. Furthermore, the film is clearly multi-oriented in plane. As indicated in the figure by thick
arrows, the majority orientation is the LCMO(110) which is grown on YSZ in a cube-on-diagonal way. The minority is
the cube-on-cube grown LCMO(100), as labelled by thinner short arrows. The very weak spots indicated by long thin
arrows are due to the LCMO(120). Obviously there exists a high density of 45$^\circ$ and 63.4$^\circ$ twin
boundaries in plane. Along LCMO(110), the lattice mismatch is -4.8\%, so the film is compressively strained. On
the contrary, along LCMO(100) the lattice mismatch is +7.3\%, if we assume that every five LCMO unit cells occupy
four YSZ unite cells, thus the film is tensilely strained. Therefore it is reasonable to conclude that a large
strain exists in the film plane, which may help to explain the very rough film surface.

\begin{figure}
\begin{center}
\includegraphics[width=0.7\columnwidth]{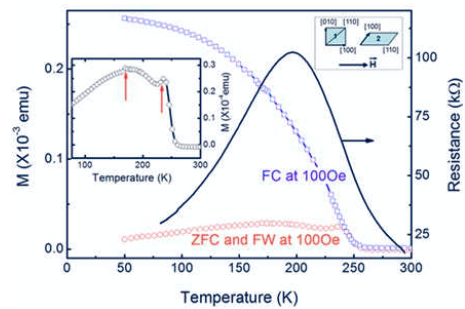}
\caption{\label{RTMT}Temperature dependence of the resistance for the LCMO thin film, and its field-cooled (FC)
and zero-field-cooled (ZFC) magnetization curves measured in a field of 100 Oe. The inset shows the two peaks on
the ZFC curve.}
\end{center}
\end{figure}

The $\rho-T$ curve of LCMO film is shown in Fig.~\ref{RTMT}. The film undergoes an insulator to metal transition
at $T_{P}\sim$~195 K, significantly lower than that of the films grown on LaAlO$_3$, which is around 260 K
\cite{Vlakhov}. The magnetization versus temperature curves, measured in a 100 Oe field applied in the film plane
roughly along Si[100], are also plotted. The field-cooled (FC) line shows a broad paramagnetic to ferromagnetic
transition with $T_C\sim$~250 K, comparable to that of the films grown on LaAlO$_3$. The drop of $T_{P}$ behind
$T_C$ in our film is due to the high-density crystallography defects (twin boundaries) in the film \cite{Blamire}.
It is worthy to note that the zero-field-cooled (ZFC) curve diverges a lot from the FC one. The magnetization
moment (\textit{M}) at 50 K under ZFC is more than 10 times lower than that under FC. Moreover, as seen in the
inset, two peaks appear on the ZFC curve, corresponding to two blocking temperatures (T$\rm _B$): a sharp one at
T$\rm _{B1}\sim 230$ K and a broad one near T$\rm _{B2}\sim 170$ K. We know that, at 0 K in zero field, the spins
in the two crystallographic structures locate in their respective [100] easy axes \cite{ODonnell}, as illustrated
at the upper-right corner of Fig.~\ref{RTMT}. In addition, the magnetostriction constant along LCMO[100] is
positive $\lambda_{100}\sim 7\times10^{-5}$ \cite{ODonnell}, so the overall in-plane tensile strain in the
direction perpendicular to the field is larger than that in the field direction, and thus the spins tend to align
perpendicular to the field. Therefore due to the magnetocrystalline anisotropy of four-fold symmetry and the
magnetoelastic anisotropy, the ZFC \textit{M} measured at low temperatures is quite small. With increase of the
temperature, the frozen spins will progressively rotate to the field direction, resulting in an increase of
\textit{M}. When the temperature is higher than $T_B$, thermal fluctuations destroy the magnetic ordering
gradually, and \textit{M} declines again. It is clear that the spins move out from LCMO[100] is more difficult for
structure \textit{1} than for structure \textit{2} due to the magnetoelastic anisotropy, therefore T$\rm _{B}$ is
higher for the former. Moreover, in the applied field, for \textit{2} the spins will gradually rotate into the
field direction, which is along its hard axes [110]. For \textit{1}, however, the spins will abruptly jump to the
field direction after they cross over [110]. Therefore T$\rm _{B1}$ is sharper than T$\rm _{B2}$. The measured ZFC
curve reveals an overall behavior of the two structures in the film.

\begin{figure}
\begin{center}
\includegraphics[width=0.8\columnwidth]{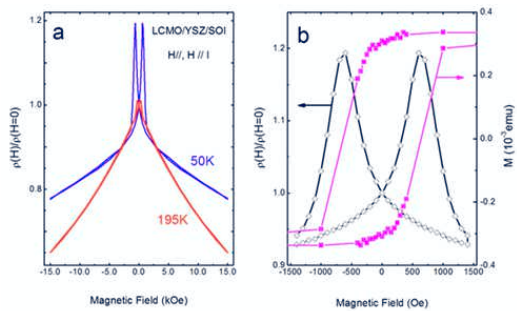}
\caption{\label{RH}(a) The resistivity ratio $\rho(H)/\rho(0)$ measured with the magnetic field and the current
all along Si[100], at T$_P$ and 50K, respectively. (b) The low field part of the 50K curve, and the magnetic
hysteresis loop measured at 50K.}
\end{center}
\end{figure}

From the investigations on microstructure and magnetization of the LCMO film, a large extrinsic LFMR based on
spin-dependent tunnelling across twin boundaries can be expected. The magnetoresistance measured in a field
scanning from -1.5 to +1.5 tesla at $T_P$ and 50 K, respectively, are shown in Fig.~\ref{RH}(a). The field was
applied roughly parallel to the measuring current, i.e. along Si[100] in the film plane. The film was cooled to
the specified temperatures in zero field. While $R(H)$ measured at 195 K shows a linear field dependence with a
small coercivity and negligible shoulders, that measured at 50 K exhibits a smaller slope at high fields, but
shows very sharp extrinsic peaks at low fields. Zoom-in of the peaks is shown in Fig.~\ref{RH}(b). For
convenience, the magnetic hysteresis loop measured at the same temperature is also plotted. It is clear that the
sample resistance peaks around $\pm$700 Oe, approximately the coercivity field, which corresponds to a state that
the magnetic spins in individual structures incline to their respective easy axes, and thus the angle between
spins in neighboring twins is the largest. A higher magnetic field can align the spins to the field direction and
reduce the resistance drastically. The peak magnitude is really impressive, which is as high as $\sim$20\%
(defined as [R(Max.)-R(0)]/R(0)). That is, in a magnetic field range 0$\sim$700 Oe, the resistivity changes for
$\sim$20\%. Up to now the only competitors were that obtained from ultrathin films \cite{Wang3}, or films with
artificial grain boundaries \cite{Mathur}. This provides an extraordinary large field sensitivity that can be
utilized for real applications. As a matter of fact, the co-existence of two in-plane orientations may be a common
feature of the manganite thin films grown on YSZ. We have noted that, Vlakhov {\textit et al.} \cite{Vlakhov} also
observed a fairly large LFMR in the LCMO thin film grown on YSZ single-crystal substrate, as can be seen clearly
in Fig. 2(b) of ref.\cite{Vlakhov}, although the underlying mechanism was not mentioned in their paper.

In summary, we have successfully deposited $\rm La_{2/3}Ca_{1/3}MnO_{3}$ thin films on YSZ-buffered SOI
substrates. The top LCMO layer was multi-domain oriented with a high density of 45$^\circ$ twin boundaries. As a
result, a quite large LFMR based on spin-dependent tunnelling was observed, which provides a most feasible
solution for real applications.




\end{document}